\documentclass{lamu}
\usepackage{graphicx}
\usepackage{epsfig}
\def\index{}
\def\Index{}
\def\d{{\rm d}}
\begin{document}

\def\emphq#1{``#1''}
\def\ssrm#1{{\textrm{\scriptsize #1}}}
\def\lambdabar{{\mathchar'26\mkern-10mu\lambda}}

\title{Measurement Theory and General Relativity}
\titlerunning{Measurement Theory and General Relativity}
\author{Bahram Mashhoon}
\institute{Department of Physics and Astronomy,
University of Missouri-Columbia, Columbia, Missouri 65211, USA}
\authorrunning{Bahram Mashhoon}
\titlerunning{Measurement Theory and General Relativity}
\maketitle

\begin{abstract}
The theory of measurement is employed to elucidate the physical basis of
general relativity.
For measurements involving phenomena with intrinsic length or time scales,
such scales must in general be negligible compared to the
(translational and rotational) scales characteristic of the motion 
of the observer.
Thus general relativity is a consistent theory of coincidences so long as
these involve classical point particles and electromagnetic rays
(geometric optics).
Wave \emphq{optics} is discussed and the limitations of the standard theory
in this regime are pointed out.
A nonlocal theory of accelerated observers is briefly described
that is consistent with observation and excludes the possibility of existence
of a fundamental scalar field in nature.
\end{abstract}

\section{Introduction}

The quantum \Index{theory of measurement}
deals with observers and measuring devices
that are all inertial.
The universality of gravitational interaction implies, however,
that gravitational fields cannot be ignored in general.
Moreover, most measurements are performed in laboratories on the Earth,
which --- among other motions --- rotates about its proper axis;
in fact, measurements are generally performed by devices and observers that
are accelerated.
It is therefore necessary to investigate the assumptions that underlie the
extension of physics to accelerated systems and gravitational fields.
This amounts to a determination of the physical foundations of Einstein's
theory of gravitation inasmuch as this theory is in agreement with all
observational data available at present \cite{ref1}.
A critical examination of general relativity from the standpoint of
measurement theory leads to certain basic 
limitations \index{general relativity!limitations}
that are the main subject of this paper.

\section{Physical Elements of General Relativity}

The basic concepts of general relativity can be uniquely determined starting 
from the consideration of what observers would measure in physical experiments.
This results in the four building blocks of general relativity that are
described below.

(i) The fundamental laws of microphysics have been formulated with
respect to \Index{inertial observers}.
The measurements of inertial observers in Minkowski spacetime are connected
via inhomogeneous Lorentz transformations (i.e. Poincar\'e transformations).
An inertial observer is an observer at rest in an inertial reference system;
in fact, such an observer can be thought of as carrying a natural 
orthonormal tetrad frame $\lambda^\mu_{(\alpha)}$ along its worldline.
Here 
$\lambda^\mu_{(0)} = \d x^\mu/\d \tau$
is the vector tangent to the worldline (\emphq{time axis}) and
$\lambda^\mu_{(i)}$, $i=1,2,3$, are the natural spatial axes
of the frame so that $\lambda^\mu_{(\alpha)} = \delta^\mu_\alpha$.
Thus Maxwell's equations in this inertial frame refer to the fields
actually measured by these standard observers, i.e. 
$F_{\mu\nu}\,\lambda^\mu_{(\alpha)}\,\lambda^\nu_{(\beta)}
\rightarrow  ({\bf E}, {\bf B})$.
One can consider other inertial observers as being at rest in other inertial
systems in uniform motion with respect to the original reference system 
described above.
To express the measurements of the other observers, one could transform to 
their rest frames; alternatively, one could consider physics in the original 
inertial system and simply describe all measurements with respect to a 
single system of inertial coordinates $x^\alpha=(ct,{\bf x})$.
In the latter case, which is adopted here for the sake of convenience, one
can describe the determination of the electromagnetic field by a moving
inertial observer as the projection of the field on the observer's frame,
\begin{equation}
  \hat F_{(\alpha)(\beta)} = F_{\mu\nu}
  \hat\lambda^\mu_{(\alpha)}   \hat\lambda^\nu_{(\beta)}.
\end{equation}
Let us now suppose that inertial observers choose to employ arbitrary 
smooth spacetime coordinates ${x'}^\mu={x'}^\mu(x^\alpha)$.
It turns out that --- so long as the observers remain inertial ---
this extension is purely mathematical in nature and can be accomplished 
without introducing any new physical assumption into the theory.
Consider, for instance, the \Index{Lorentz force law} for a particle of 
mass $m$ and charge $q$,
\begin{equation}
  m {\d^2 x^\mu\over \d \tau^2} =
  q F^\mu{}_\nu {\d x^\nu\over \d \tau}.
\label{equ2}
\end{equation}
Here $\d\tau$ is the invariant spacetime interval measured along the path of
the particle by the standard inertial observers, i.e. 
$\d\tau=c\,\d t/\gamma$ and $\gamma$ is the Lorentz factor.
Assuming the invariance of this interval under the change of coordinates,
$\d\tau^2=\eta_{\mu\nu} \d x^\mu \d x^\nu=
g'_{\alpha\beta} \d {x'}^\alpha \d {x'}^\beta$
with 
\begin{equation}
  g'_{\alpha\beta} = \eta_{\mu\nu} 
  {\partial x^\mu\over \partial {x'}^\alpha}
  {\partial x^\nu\over \partial {x'}^\beta},
\end{equation}
one can simply write equation (\ref{equ2}) as
\begin{equation}
  m\left[ {\d^2 {x'}^\rho\over \d \tau^2} +
  {\Gamma}^{\prime\rho}_{\alpha\beta} (x') 
  {\d{x'}^\alpha\over \d\tau}
  {\d{x'}^\beta\over \d\tau}
  \right] = q {F'}^\rho{}_\sigma   {\d {x'}^\sigma\over \d \tau},
\end{equation}
with the Christoffel connection
\begin{equation}
  {\Gamma}^{\prime\rho}_{\alpha\beta} = 
  {\partial^2 x^\mu\over\partial{x'}^\alpha\partial{x'}^\beta}
  {\partial {x'}^\rho\over\partial x^\mu},
\end{equation}
and the auxiliary field variables
\begin{equation}
  {F'}^{\rho\sigma} (x') = 
  {\partial {x'}^\rho\over\partial x^\mu}
  {\partial {x'}^\sigma\over\partial x^\nu}
  F^{\mu\nu}(x).
\end{equation}
In Euclidean space, one can always introduce curvilinear
coordinates for the sake of convenience; similarly, one can introduce
arbitrary (smooth and admissible) coordinates in Minkowski spacetime.
In this way, tensors under the inhomogeneous Lorentz group become tensors
under general coordinate transformations. 

\begin{figure}[htb]
  \begin{center}
    \leavevmode
    \scalebox{0.9}{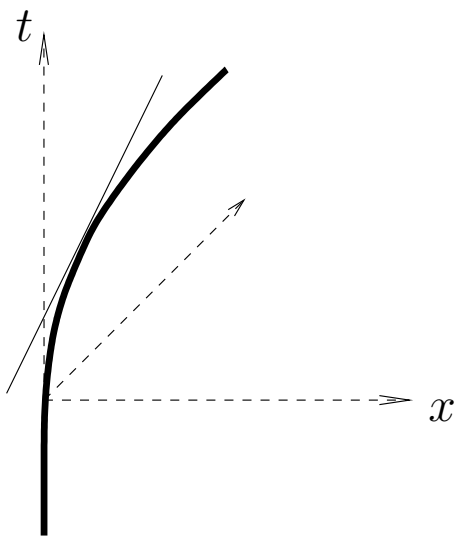}
  \end{center}
    \caption{The worldline of an accelerated observer in Minkowski spacetime
             is curved. The hypothesis of locality postulates that the
             observer is at each moment locally inertial.}
    \label{fig:1}
\end{figure}

(ii) To extend measurements to accelerated observers\index{accelerated observer} 
in Minkowski spacetime, a physical hypothesis is required that would connect
the measurement of accelerated and inertial observers.
In the standard approach to the theory of relativity, the assumption is that an 
accelerated observer is at each instant physically equivalent to a 
hypothetical momentarily comoving inertial observer.
Thus an accelerated observer passes through an infinite sequence of such 
hypothetical inertial observers.
Mathematically, this basic assumption is equivalent to replacing a curve
by its tangent vector at each point as illustrated in Figure \ref{fig:1}.
This assumption is clearly valid for Newtonian point particles, since at 
each instant the accelerated particle and the momentarily comoving inertial
particle have the same state, i.e. the same position and velocity.
Moreover, it can be naturally extended to all pointlike phenomena; that is, the
assumption is also valid if all phenomena are thought of in terms of pointlike
\emph{coincidences} of Newtonian point particles and null rays.
However, in more general cases involving intrinsic temporal and spatial scales 
the above assumption will be referred to as \emphq{the hypothesis of locality}
\cite{ref2}.
Imagine, for instance, an accelerated measuring device; clearly, it is affected
by internal inertial effects.
If these inertial effects integrate to a perceptible influence on the outcome
of a measurement, the hypothesis of locality is violated.
On the other hand, if the timescale of the measurement is so short that the 
influence of the inertial effects is negligible, then the device is 
\emphq{standard}, i.e. its acceleration can be locally ignored.
The hypothesis of locality applied to a clock implies that a standard clock 
will measure \Index{proper time} $\tau$ along its path;
therefore, the hypothesis of locality is the generalization of the
\emphq{clock hypothesis} to all standard measuring 
devices \cite{ref3,ref4,ref5,ref6,ref7}.
Moreover, the local equivalence of an accelerated observer with an infinite
sequence of comoving inertial observers endows the accelerated observer with the
continuously varying tetrad system of the inertial observers.
This variation can be characterized by a translational acceleration 
${\bf g}(\tau)$ and a rotation of the spatial frame with frequency 
${\bf \Omega}(\tau)$; 
alternatively, one may associate acceleration scales 
(such as $c^2/g$ and $c/\Omega$)
with the motion of the observer \cite{ref8,ref9}.

The extension of measurements to all observers that can use arbitrary
coordinates in Minkowski spacetime implies that one can formulate physical
laws in a \emph{generally covariant} form.
To extend this covariance further to curved spacetime manifolds, Einstein's
principle of equivalence\index{equivalence principle} 
is indispensable.

(iii) Einstein's principle of equivalence
embodies the universality of the
gravitational interaction and is the cornerstone of general relativity.
This principle generalizes a result of Newtonian gravitation that is directly
based upon the principle of equivalence of inertial and gravitational masses.
Einstein postulated a certain equivalence between an observer in a
gravitational field and an accelerated observer in Minkowski spacetime.
This heuristic principle, when combined with the hypothesis of locality,
implies that an observer in a gravitational field is locally inertial.
Thus gravitation has to do with the way local inertial frames are connected
to each other.
The simplest possibility is through the pseudo-Riemannian curvature of the
spacetime manifold;
therefore, in general relativity the gravitational field is identified with
the spacetime curvature.

(iv) The correspondence between general relativity and Newton's theory of
gravitation is established via the gravitational field equation.
That is, within the framework of Riemannian geometry the gravitational 
field equations are the simplest generalizations of Poisson's equation,
$\nabla^2\Phi_\ssrm{N}=-4\pi G\rho$, for the Newtonian potential
$\Phi_\ssrm{N}$.
In general relativity, the Newtonian potential is generalized and 
replaced by the ten components of the metric tensor $g_{\mu\nu}$;
similarly, the acceleration of gravity is replaced by the Christoffel
connection $\Gamma^\mu_{\alpha\beta}$ and the tidal matrix
$\partial^2 \Phi_{\ssrm{N}}/\partial x^i \partial x^j$ is replaced by the
Riemann curvature tensor $R_{\mu\nu\rho\sigma}$.
In Newtonian gravitation, the trace of the tidal matrix is connected to the
local density of matter $\rho$ by Newton's constant of gravitation.
Similarly, in general relativity the trace of the Riemann tensor is 
connected to the energy-momentum tensor of matter,
\begin{equation}
  \label{equ:7}
  R_{\mu\nu} - {1\over 2} g_{\mu\nu} g^{\alpha\beta} R_{\alpha\beta}= 
  {8\pi G\over c^4} T_{\mu\nu}.
\end{equation}

\section{Measurements of Accelerated Observers}

The primary measurements of an observer are those of duration and distance.
In general relativity, the hypothesis of locality is indispensable for
the interpretation of the results of measurements by accelerated observers.
In particular, we define \emphq{standard} measuring devices
to be those that are compatible with the locality assumption.
Thus a standard clock measures \Index{proper time} along its trajectory;
similarly, a standard measuring rod is usually assumed to provide a proper
measure of distance.
At each instant of time, the accelerated observer is momentarily equivalent
to a hypothetical comoving inertial observer;
therefore, both observers have the instantaneous Euclidean space in common.
It would appear then that placing standard measuring rods one next to the other
and so on should lead to the proper measurement of spatial distances
by accelerated observers.

An important issue is the extent to which such measurements of time and
distance can lead to the establishment of an admissible coordinate system
around the accelerated observer.
Well-known investigations have led to the result that such coordinate systems
have limited spatial extent given by the acceleration lengths
(e.g. $c^2/g$ and $c/\Omega$), since these are the only length scales
in the problem.
The method of construction of accelerated coordinate systems could even be
nonlocal; however, limitations would still exist as recently pointed out by 
Marzlin \cite{ref10}.
It might therefore appear that (local and nonlocal) coordinate systems could
in general be constructed in a cylindrical region around the worldline
of the accelerated observer.
However, this conclusion is ultimately based upon the use of standard measuring
rods whose existence turns out to be in conflict with the hypothesis of
locality.
A fundamental problem associated with length measurements is the following:
a standard measuring rod, however small, has nevertheless a nonzero spatial
extent whereas the hypothesis of locality is only pointwise valid.
This implies a rather basic limitation on the measurement of length by
accelerated observers and can be illustrated by the following thought 
experiment.
Imagine two observers $O_1$ and $O_2$ at rest in an inertial frame.
For $t\leq 0$, their coordinates 
$x^\alpha=(ct,{\bf x})$ are $(ct,0,0,0)$ and $(ct,L,0,0)$, respectively.
At $t=0$, they are accelerated from rest along the $x$-direction
\emph{in exactly the same way} so that at time $t>0$ each has a velocity
${\bf v} = v {\bf \hat x}$.
The distance between $O_1$ and $O_2$ as measured by observers at rest in the
inertial frame is always $L$, since 
\begin{equation}
  x_1(t) = \int_0^tv\,\d t \qquad \textrm{and} \qquad
  x_2(t)=L+\int_0^tv\,\d t
\end{equation}
for $t>0$ and $x_2(t)-x_1(t)=L$.
What is the distance between $O_1$ and $O_2$ as measured by comoving observers?
It turns out that the hypothesis of locality provides a unique answer
to this question only in the limit $L\rightarrow 0$.
To show this, let us first note that at a given time
$\hat t>0$, $O_1$ and $O_2$ have the same speed $\hat v=c\beta$.
The hypothesis of locality implies that the accelerated observers pass 
through an infinite sequence of momentarily comoving inertial observers.
Thus imagine the Lorentz transformation between the inertial frame
$x^\alpha=(ct,{{\bf x}})$ and the \emphq{instantaneous} inertial rest frame
${x'}^\alpha = (ct',{\bf x'})$ of the observers at $\hat t$ given by
\begin{equation}
c(t-\hat t\,)=\gamma(ct'+\beta x'), \quad 
x-\hat x=\gamma (x'+c\beta t'), \quad
y=y', \quad z=z',
\label{equ:9}
\end{equation}
where $\gamma=(1-\beta^2)^{-1/2}$ is the Lorentz factor at $\hat t$.
The events with coordinates 
$O_1: (c\hat t, x_1,0,0)$ and
$O_2: (c\hat t, x_2,0,0)$ in the original inertial frame have
coordinates $O_1: (c{t'}_1, x'_1,0,0)$
and $O_2: (c{t'}_2, x'_2,0,0)$ in the instantaneous inertial frame.
It follows from the Lorentz transformation (\ref{equ:9}) that
$L'=x'_2-x'_1=\gamma L$.
This has a simple physical interpretation: The Lorentz-FitzGerald
contracted distance between $O_1$ and $O_2$ is always $L$, hence the
\emphq{actual} distance between $O_1$ and $O_2$ must be larger by the 
Lorentz $\gamma$-factor.
One can imagine that the distance between $O_1$ and $O_2$ is populated by a 
large number of hypothetical accelerated observers moving in exactly the same
way as $O_1$ and $O_2$ and carrying infinitesimal measuring rods that are
placed side by side to measure the distance under consideration.

It must be equally correct to replace the infinite sequence of inertial systems
${x'}^\alpha = (ct', {\bf x'})$ by a continuously moving frame.
To this end, we must choose the worldline
$\overline{x}^\mu(\tau)$ of one of the accelerated observers
--- such as $O_1$, $O_2$, or any of the hypothetical observers in between 
the two ---
and note that at any instant of proper time
$\tau$ along the worldline, this
fiducial observer is in a Euclidean space with Cartesian coordinates 
${\bf X}$ in accordance with the hypothesis of locality.
The connection between the coordinates $x^\mu$ in the original inertial
frame and the new coordinates $X^\mu$ is given by $X^0=\tau$ and
\begin{equation}
  x^\mu = \overline{x}^\mu(X^0)+X^i\,\overline{\lambda}^\mu_{(i)},
\end{equation}
where $\overline{\lambda}^\mu_{(i)}$ is the natural tetrad frame along the
worldline of the reference observer.
Specifically, the fiducial observer is instantaneously inertial by the
hypothesis of locality and hence assigns coordinates $X^0=\tau$
and $X^i=\sigma \xi_\mu \overline{\lambda}^\mu_{(i)}$ to spacetime
events.
Here $\xi^\mu$ is a unit spacelike vector normal to 
$\overline{\lambda}^\mu_{(0)}$ at $\overline{x}^\mu(\tau)$ along
a straight line that connects $\overline{x}^\mu(\tau)$ to an event with
coordinates $x^\mu$  in the original background inertial frame,
$\xi_\mu \overline{\lambda}^\mu_{(i)}$ are direction cosines and
$\sigma=\left| {\bf X} \right|$ is the proper length of this
spacelike line segment.
To develop this approach further, it is necessary to specify the motion 
explicitly.
Thus we assume that $O_1$ and $O_2$ are uniformly accelerated with 
acceleration $g$ and we choose $O_1$ to be the fiducial observer.
The natural orthonormal nonrotating tetrad frame along the worldline
of $O_1$ is given by
\begin{eqnarray}
  \overline{\lambda}^\mu_{(0)} &=& (\gamma,\beta\gamma,0,0),\\
  \overline{\lambda}^\mu_{(1)} &=& (\beta\gamma,\gamma,0,0),\\
  \overline{\lambda}^\mu_{(2)} &=& (0,0,1,0),\\
  \overline{\lambda}^\mu_{(3)} &=& (0,0,0,1),
\end{eqnarray}
just as for the Lorentz transformation (\ref{equ:9}).
Then the inertial frame $x^\alpha=$ $(ct,x,y,z)$ and the Fermi frame
$X^\alpha=(cT,X,Y,Z)$ are connected by
\begin{eqnarray}
  ct &=& \left(X+{c^2\over g}\right) \sinh \left(gT/c\right),\label{equ:15}\\
  x  &=& \left(X+{c^2\over g}\right) \cosh \left(gT/c\right)-{c^2\over g},
\label{equ:16}
\end{eqnarray}
$y=Y$ and $z=Z$.
The spatial origin of the new coordinate system is occupied by $O_1$ such 
that $\overline{x}^\mu(\tau)={\cal L} (\beta \gamma,\gamma-1,0,0)$,
where $\beta=\tanh(\tau_1/{\cal L})$,
$\gamma=\cosh(\tau_1/{\cal L})$, ${\cal L}=c^2/g$ is the acceleration length
and $\tau_1$ is the proper time along $O_1$.
As before, at any given time $\hat t>0$ the events 
$O_1: (c\hat t,x_1,0,0)$ and
$O_2: (c\hat t,x_2,0,0)$ now correspond to 
$O_1: (\tau_1, X_1,0,0)$ and
$O_2: (\tau_2,X_2,0,0)$, where $x_2-x_1=L$ and $X_1=0$ by construction.
The distance between $O_1$ and $O_2$ in this Fermi frame is then given by
$L_{\ssrm{F}} = X_2 - X_1 = X_2$.
It follows from equations (\ref{equ:15}) and (\ref{equ:16}) that
\begin{eqnarray}
  c \hat t &=& {\cal L} \sinh (\tau_1/{\cal L}),\label{equ:17}\\
  x_1      &=& {\cal L} \left[\cosh (\tau_1/{\cal L})-1\right],\label{equ:18}\\
  c \hat t &=& (X_2+{\cal L}) \sinh (\tau_2/{\cal L}),\label{equ:19}\\
  x_2      &=& (X_2+{\cal L}) \cosh (\tau_2/{\cal L})-{\cal L}.\label{equ:20}
\end{eqnarray}
Equations (\ref{equ:19}) and (\ref{equ:20}) can be written as
\begin{equation}
  (X_2+{\cal L})^2 = (x_2+{\cal L})^2 - c^2 {\hat t}^2,
\end{equation}
where $x_2=x_1+L$ and $x_1$ and $\hat t$ are given by
equations (\ref{equ:18}) and (\ref{equ:17}), respectively.
Thus one finds that 
\begin{equation}
  L_{\ssrm{F}} = {\cal L} \left[ (1+2\epsilon\gamma+\epsilon^2)^{1/2}
    -1\right],
\end{equation}
where $\epsilon=L/{\cal L}=gL/c^2$ and
$\gamma=(1+g^2{\hat t}^2/c^2)^{1/2}$.
The length in the Fermi frame $L_{\ssrm{F}}$ must be compared with 
the corresponding result from the instantaneous Lorentz frame 
$L'=\gamma L$; indeed, the ratio $L_{\ssrm{F}}/L'$ approaches unity only
in the limit $\epsilon\rightarrow 0$.
This is a remarkable result that has far-reaching consequences.
Let us note that for $\epsilon \ll 1$,
\begin{equation}
  L_{\ssrm{F}}/L' \approx 1-{1\over 2} \beta^2 \gamma\epsilon
\end{equation}
to first order in $\epsilon$;
however, over a long time $\gg c/g$ the quantity $\beta^2\gamma\epsilon$ 
may not remain small compared to unity.
Moreover, $L_{\ssrm{F}}/L'\rightarrow0$ as
$g\hat t/c\rightarrow \infty$ and hence $\gamma\rightarrow\infty$.
It follows from these considerations that consistency is achieved for
$\gamma\epsilon \rightarrow 0$;
hence, the acceleration length and time, i.e. $c^2/g$ and $c/g$,
respectively, place severe limitations on the domain of applicability of the
hypothesis of locality.
Furthermore, let us suppose that the Fermi frame is established along $O_2$
instead of $O_1$. 
Then the resulting distance would be different from $L_{\ssrm{F}}$;
however, all such lengths agree in the $\epsilon \rightarrow0$ limit.

It is interesting to mention here another measure of distance from
$O_1$ and $O_2$ using light signals.
Let $O_1$ send a signal at $\tau_1^-$ that reaches $O_2$ at $\tau_2$ and
is immediately returned to $O_1$.
The return signal reaches $O_1$ at $\tau_1^+$, where 
$\tau_2=(\tau_1^- + \tau_1^+)/2$.
Observer $O_1$ would then determine the distance to $O_2$ via 
$L_{\ssrm{ph}}=c(\tau_1^+-\tau_1^-)/2$, which works out to be
\begin{equation}
  L_{\ssrm{ph}}={\cal L}\,\ln{(1+L_{\ssrm{F}}/{\cal L})}.
  \label{equ:24} 
\end{equation}
It is clear by symmetry that if $O_2$ initiates a light signal to $O_1$, etc.,
then the resulting light travel time would be different, since in 
equation (\ref{equ:24}) the Fermi length would be the one determined on the
basis of $O_2$ as the fiducial observer.
Nevertheless, for $\gamma\epsilon$ negligibly small all these length
measurements agree with each other.

The simple example that has been worked out here can be generalized to
arbitrary but identical velocity for $O_1$ and $O_2$.
The comparison of the instantaneous local inertial frame with the 
continuously moving geodesic frame leads to the conclusion that the basic
length and time scales under consideration must in general be negligible 
compared to the relevant acceleration scales.
This has significant consequences for the comparison of theory and experiment
in general relativity \cite{ref2};
in particular, the physical significance of Fermi coordinates is in general
further limited to the immediate neighborhood of the observer and wave
equations are meaningful only within this domain.

It follows from these considerations that the physical dimensions of any standard
measuring device must be negligible compared to the relevant acceleration
length ${\cal L}$ and the duration of the measurement must in general be
negligible compared to ${\cal L}/c$.
These are not significant limitations for typical accelerations in the
laboratory; for instance, for the Earth's acceleration of gravity
$c^2/g\simeq 1\,$lyr.
Moreover, observers at rest on the Earth typically refer their measurements
to rotating Earth-based coordinates;
hence, this coordinate system is mathematically valid up to a 
\emphq{light cylinder} at a radius of
${\cal L}=c/\Omega \simeq 28\,$AU.
But physically valid length measurements can extend over a neighborhood of the
observer with a radius much smaller than ${\cal L}$.
In fact, this \emphq{light cylinder} has no bearing on astronomical observations,
since observers simply take into account the absolute rotation of the Earth and
reduce astronomical data by taking due account of aberration and Doppler
effects.

The standard \emphq{classical} measuring device of mass $\mu$ has wave
characteristics, given by its Compton wavelength $\hbar/\mu c$
and period $\hbar/\mu c^2$, that must be negligible in comparison with the 
scales of length and time that characterize the device as a consequence of the
quasi-classical approximation.
For instance, a clock of mass $\mu$ must have a resolution exceeding
$\hbar/\mu c^2$; similarly, the mass of a clock with resolution $\theta$
must exceed $\hbar /\theta c^2$.
These assertions follow from the application of the uncertainty principle
to measurements performed by a standard device \cite{ref11,ref12}.
When such quantum limitations are combined with the classical limitations
discussed above, on finds that ${\cal L} \gg \hbar/\mu c$;
therefore, the translational acceleration of a standard classical measuring
device must be much less than $\mu c^3/\hbar$ and its rotational frequency
must be much less than $\mu c^2/\hbar$.
The idea of the existence of a maximal proper acceleration is due to
Caianiello \cite{ref13,ref2,ref14}.

\section{Measurements in Gravitational Fields}

The physical results of the previous section can be extended to local
measurements in a gravitational field via an interpretation of the Einstein 
principle of equivalence\index{equivalence principle|(} 
in terms of the gravitational Larmor theorem.
\index{Larmor theorem!gravitational|(}

A century ago, Larmor \cite{ref15} established a local equivalence
between magnetism and rotation for all particles with the same charge to mass
ratio $(q/m)$.
That is, charged particle phenomena in a magnetic field correspond to those 
in a frame rotating with the Larmor frequency 
${\bf \Omega}_{\ssrm{L}} = q {\bf B}/2mc$.
This local relation is valid to first order in field strength for slowly 
varying fields and slowly moving charged particles.
Such a correspondence also exists for electric fields and linearly accelerated 
frames.
It turns out that Larmor's theorem can be generalized in a natural way to
the case of gravitational fields.

The close analogy between Coulomb's law of electricity and Newton's law of
gravitation leads to an interpretation of Newtonian gravity in terms of
nonrelativistic theory of the gravitoelectric field.
Moreover, any theory that combines Newtonian gravity with Lorentz invariance
in a consistent manner is expected to contain a gravitomagnetic field as well.
In fact, in general relativity the exterior spacetime metric for a rotating
mass may be expressed in the linear approximation as
\begin{equation}
  \d s^2 = -c^2(1-{2\over c^2}\Phi_{\ssrm{N}}) \,\d t^2
  +(1+{2\over c^2}\Phi_{\ssrm{N}}) \,\delta_{ij} \,\d x^i\,\d x^j
  - {4\over c} ({\bf A}_g\cdot\d {\bf x}) \,\d t,
\end{equation}
where $\Phi_{\ssrm{N}} = GM/r$ is the Newtonian potential and
${\bf A}_g=G {\bf J}\times{\bf r}/cr^3$
is the gravitomagnetic vector potential.
The gravitoelectric and gravitomagnetic fields are then given by
${\bf E}_g=-{\bf \nabla} \Phi_{\ssrm{N}}$
and ${\bf B}_g = {\bf \nabla}\times{\bf A}_g$, respectively.

It is possible to formulate a gravitational Larmor theorem \cite{ref16}
by postulating that the gravitoelectric and gravitomagnetic charges are given
by $q_E=-m$ and $q_B=-2m$, respectively.
In fact, $q_B/q_E=2$ since gravitation is a spin-2 field.
Thus ${\bf \Omega}_{\ssrm{L}}=-{\bf B}_g/c$, which is consistent 
with the fact that an ideal gyroscope at a given position in space would
precess in the gravitomagnetic field with a frequency 
${\bf \Omega}_{\ssrm{P}}={\bf B}_g/c$.
The general form of the gravitational Larmor theorem
\index{Larmor theorem!gravitational|)}
 is then an interpretation
of the Einstein principle of equivalence\index{equivalence principle|)}
for linear gravitational fields in a
finite neighborhood of an observer;
for instance, in the gravitational field of the Earth an observer can be
approximately inertial within the \emphq{Einstein elevator} if the ``elevator''
falls freely with acceleration 
$g\sim GM/r^2$ while rotating with frequency $\Omega\sim GJ/c^2r^3$.
It follows that the relevant gravitoelectromagnetic acceleration lengths are
given by $c^2/g$ and $c/\Omega$ in this case
and the restrictions discussed in the previous
section would then apply to the measurements of an observer in a gravitational
field as well.
These limitations are generally expected to be important for the
post-Newtonian corrections of high order in relativistic gravitational systems.

General relativity has found applications mostly in astronomical systems,
where Newtonian results have been extended to the relativistic domain.
In particular, small post-Newtonian corrections are usually included in the
equations of motion.
Suppose, for instance, that one is interested in the distance between the
members of a relativistic binary system.
It follows from our considerations that such a length
--- which corresponds in the Newtonian theory to the Euclidean distance ---
may not be well defined.
However, the resulting discrepancy could be masked by other parameters; 
that is, this circumstance may be difficult to ascertain experimentally
since the comparison of data with the theory generally involves
parameters that are not independently available and whose particular
values need to be determined from the data.

Let us next consider tidal accelerations within the 
\emphq{Einstein elevator}.
For a device of dimension $\hat \delta$, the tidal acceleration $\hat g$ 
is given by the Jacobi equation and can be estimated by 
$\hat g\sim K\hat\delta$, where $K$ is a typical component of the tidal matrix 
$K_{ij} =c^2 R_{\mu\nu\rho\sigma} \lambda^\mu_{(0)}\lambda^\nu_{(i)}
\lambda^\rho_{(0)}\lambda^\sigma_{(j)}$.
According to the results of the previous section 
$\hat g\ll \mu c^3/\hbar$, where $\mu$ is the mass of the device.
Imagine, for instance, such a device on a star of mass $M$ and radius $R$
that is undergoing \emphq{complete} spherical gravitational collapse.
In this case , $K\sim GM/R^3$ and
$\hat\delta\ll c^2/\hat g$ imply that 
${\hat \delta}^2\ll c^2R^3/GM$.
On the other hand, the requirements that $\mu\ll M$ and 
$\hat\delta\gg\hbar/\mu c$ result in
\begin{equation}
  R^3 \gg {GM\over c^2} \left({\hbar\over Mc}\right)^2 = 
  {\hbar\over Mc} L_{\ssrm{P}}^2,
\end{equation}
where  $L_{\ssrm{P}}=(\hbar G/c^3)^{1/2}$ is the Planck length 
($\simeq 10^{-33}\,$cm) that is the geometric mean of the gravitational radius
$GM/c^2$ and the Compton wavelength $\hbar/Mc$
for any physical system \cite{ref17}.
Thus collapse to a classical point singularity is meaningless on the basis of
these considerations.

\section{Wave Phenomena}

\begin{figure}[htb]
  \begin{center}
    \leavevmode
    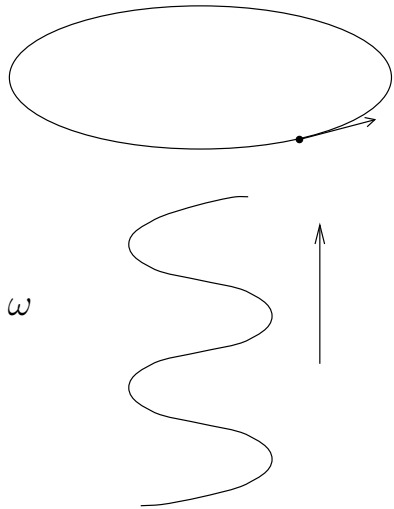
    \caption{A thought experiment involving the measurement of the 
      frequency of a normally incident plane monochromatic electromagnetic
      wave of frequency $\omega$ by a uniformly rotating observer.}
    \label{fig:2}
  \end{center}
\end{figure}

Classical waves have intrinsic scales and are thus expected to be in conflict
with the hypothesis of locality;
indeed, for an electromagnetic wave of (reduced) wavelength $\lambdabar$
the expected deviation from the hypothesis of locality is expected to be
of the form $\lambdabar/{\cal L}$.
More specifically, let us consider the problem of determination of the period 
of an incident electromagnetic wave by an accelerated observer.
The observer needs to measure at least a few oscillations of the wave before
a reasonable determination of the period can be made;
therefore, the curvature of the observer's worldline cannot be neglected unless
$\lambdabar/{\cal L}$ is too small to be observationally significant.
It follows that the instantaneous Doppler
and aberration formulas are in
general valid only in the eikonal limit 
$\lambdabar/{\cal L}\rightarrow0$.
The issues involved here can be illustrated by a simple thought experiment.
Let us consider an observer rotating with uniform speed $c\beta$ and
frequency $\Omega$ in the positive sense around the origin on a circle
of radius $r=c\beta/\Omega$ in the $(x,y)$-plane.
A plane electromagnetic wave of frequency $\omega$ is incident along the 
$z$-axis and the rotating observer measures its frequency.
According to the hypothesis of locality, the observer is at each instant
momentarily inertial and hence $\omega'=\gamma\omega$ according to the 
transverse Doppler effect.\index{Doppler!effect}
This is illustrated in Figure \ref{fig:2}.
On the other hand, if we assume that the hypothesis of locality applies to the
field measurement,
\begin{equation}
  F_{(\alpha)(\beta)} (\tau) = F_{\mu\nu} \lambda^\mu_{(\alpha)}
  \lambda^\nu_{(\beta)},
  \label{equ:27}
\end{equation}
and the instan\-taneously determined electro\-magnetic field 
$F_{(\alpha)(\beta)}(\tau)$ is then Fou\-rier
ana\-lyzed over proper time ---
which is definitely a nonlocal procedure --- to determine its frequency content,
then we find that 
$\omega'=\gamma(\omega\mp \Omega)$.
Thus $\omega'=\gamma\omega (1\mp \lambdabar/{\cal L})$, where 
${\cal L}=c/\Omega$; hence, the instantaneous Doppler result is recovered 
for $\lambdabar\rightarrow 0$.
The upper (lower) sign here refers to right (left) circularly polarized
incident wave.
Apart from the Lorentz factor $\gamma$ that refers to the time dilation
involved here, the result for $\omega'$ has a simple physical interpretation:
The electromagnetic field rotates with frequency $\omega$ ($-\omega$)
about the $z$-axis for an incident right (left) circularly polarized wave, so 
that the field rotates with respect to the observer with frequency 
$\omega-\Omega$ ($-\omega-\Omega$).
Thus the helicity of the radiation couples to the rotation of the observer,
i.e.
$\hbar\omega'=\gamma(\hbar\omega - {\bf s}\cdot{\bf \Omega})$;
in fact, this is an example of the general phenomenon of spin-rotation
coupling \cite{ref18,ref19,ref20,ref21,ref22,ref23,ref24}.
For instance, for experiments on the Earth the \emphq{nonrelativistic}
Hamiltonian for a spin-${1\over2}$ particle should be supplemented by
\begin{equation}
  {\cal H}_{\ssrm{SR}} = -{\bf s}\cdot{\bf\Omega} +
  {\bf s}\cdot{\bf \Omega}_{\ssrm{P}},
  \label{equ:28}
\end{equation}
where ${\bf \Omega}$ is the frequency of Earth's rotation and 
${\bf \Omega}_{\ssrm{P}}$ is the gravitomagnetic precession frequency.
The second term in equation (\ref{equ:28}) illustrates the gravitational
Larmor theorem.\index{Larmor theorem!gravitational}
It is interesting to note that 
$\hbar\Omega\sim 10^{-19}\,$eV and $\hbar\Omega_{\ssrm{P}}\sim 10^{-29}\,$eV;
in fact, recent experiments \cite{ref25,ref26} have demonstrated the existence
of the first term in (\ref{equ:28}).
Moreover, the position dependence of the second term in (\ref{equ:28})
indicates the existence of a gravitomagnetic Stern-Gerlach force 
$-\nabla({\bf s}\cdot{\bf \Omega}_{\ssrm{P}})$ that is purely
spin dependent and violates the universality of free fall.
For instance, neutrons in different spin states in general fall differently 
in the gravitational field of a rotating mass;
similarly, the gravitational deflection of polarized light is affected by the
rotation of the mass.
That is, in addition to, and about, the Einstein deflection angle
$\Delta=4GM/c^2D$, there is a splitting due to the helicity-rotation 
coupling by a much smaller angle $\delta=4\lambdabar GJ/c^3D^3$, where $D$ is
the impact parameter 
for radiation propagating normal to the rotation axis and over a pole of
the rotating mass \cite{ref18,ref16}.
As $\lambdabar/{\cal L}\rightarrow0$, $\delta\rightarrow 0$ and 
hence the standard result for a null geodesic is recovered.
 
To explain all of the experimental tests of general relativity, it is
sufficient to consider all wave phenomena only in the JWKB limit.
That is, geometric \emphq{optics} is all that is required;
no gravitational effect involving wave \emphq{optics} has ever been detected
thus far.
An interesting opportunity for detecting such effects would come about if the
quasinormal modes (QNMs) of black holes could be observed.
The infinite set of QNMs corresponds to damped oscillations of a black hole
that come about as the black hole divests itself of the energy of the 
external perturbation and returns to a stationary state;
therefore, these ringing modes of black holes appear as
${\cal A} \exp(-i\omega t)$ at late times far from a black hole.
Here ${\cal A}$ is the amplitude of the oscillation that depends on the
strength of the perturbation as well as the black hole response,
while $\omega=\omega_0 -i\Gamma$ with $\Gamma\geq 0$ is purely a function 
of mass $M$, angular momentum $J$ and charge $Q$ of the black hole, i.e. 
$\omega=\omega_{jmn}(M,J,Q)$, where $j$, $m$ and $n$ are parameters
characterizing the total angular momentum of the radiation field, its
component along the $z$-axis and the mode number, respectively \cite{ref27}.
The mode number $n=0,1,2,\dots$, generally refers to the fundamental,
first excited state, etc., of the perturbed black hole with $j$ and $m$; 
in fact, $\Gamma$ increases with $n$ so that the higher excited states are more
strongly damped.
The fundamental least-damped gravitational mode with $j=2$ and $n=0$ for a 
Schwarzschild black hole is given by
\begin{eqnarray}
  \omega_0/2\pi &\approx& 10^4 (M_\odot /M)\;\textrm{Hz},\label{equ:29}\\
\nonumber\\
  \Gamma^{-1} &\approx& 6\times 10^{-5} (M/M_\odot) \;\textrm{sec},
  \label{equ:30}
\end{eqnarray}
so that even this mode is rather highly damped and would therefore be very
difficult to observe.
The damping problem improves by an order of magnitude if the black hole 
rotates rapidly;
however, the observational difficulties would still be considerable.
The observation of such a mode would be very significant physically 
since, among other things, near an oscillating black hole
${\cal L}_g\sim GM/c^2$ and with $\lambdabar=c/\omega_0$,
 we have $\lambdabar/{\cal L}_g\sim 1$, so that wave
\emphq{optics} can be explored in the gravitational field of a black hole.

It is necessary to examine the justification for the local field assumption
(\ref{equ:27}), since it leads --- in the thought experiment of Figure
\ref{fig:2} --- to the result that a normally incident right circularly 
polarized wave with $\omega=\Omega$ would stand completely still with 
respect to the observer.
This circumstance is in contradiction with expectations based on elementary 
notions of relativity theory \cite{ref28}. 
In fact, at $\omega=\Omega$ one has $\lambdabar/{\cal L}=1$ and it is 
possible to argue that the hypothesis of locality must be violated.
To this end, imagine an accelerated charged particle in the nonrelativistic 
approximation.
The particle radiates electromagnetic waves with characteristic wavelength
$\lambdabar\sim{\cal L}$;
therefore, it is expected that such a particle would not be locally inertial
and that (\ref{equ:27}) is violated.
Indeed, the equation of motion of the particle is given by 
\begin{eqnarray}
  m {\d^2{\bf x}\over\d t^2} -
  {2\over 3} {q^2\over c^3} {\d^3{\bf x}\over \d t^3} + \dots 
  = {\bf f}.
\end{eqnarray}
The radiation reaction term --- due originally to Abraham and Lorentz ---
ensures that the particle is not pointwise inertial, since its position and
velocity are not sufficient to determine the state of the radiating particle.

These classical considerations must naturally extend to the quantum domain
as well, since quantum theory is based on the notion of wave-particle 
duality.
That is, we expect that the hypothesis of locality would be violated in the
quantum regime.
Consider, for instance, the determination of muon lifetime by Bailey
\emph{et al.} \cite{ref29} involving muons (in a storage ring at CERN) 
undergoing centripetal acceleration of 
$g=\gamma^2v^2/r\simeq 10^{21}\;\textrm{cm}\,\textrm{sec}^{-2}$.
If $\tau_\mu^0$ is the lifetime of the muon at rest, then the hypothesis of
locality would imply that the lifetime in the storage ring would be
$\tau_\mu=\gamma \tau_\mu^0$.
In the experiment, $r\simeq 7\;$m, $\gamma\simeq29$ and time dilation is 
verified at the level of $\sim10^{-3}$.
On the other hand, the deviation from the hypothesis of locality is expected 
to be of the form $\lambdabar/{\cal L}\sim 10^{-13}$, where 
$\lambdabar=\hbar/mc$ is the Compton wavelength of the muon and 
${\cal L}=c^2/g\simeq1\;$cm is the translational acceleration length.
But the functional form of this deviation is not specified by our general
intuitive considerations.
In any case, $\lambdabar/{\cal L}$ is about ten orders of magnitude 
below the level of experimental accuracy \cite{ref29}.
In fact, the decay of the muon has been considered in this case by 
Straumann and Eisele by replacing the accelerated muon by the stationary 
state of a muon in a Landau level with very high quantum number \cite{ref30}.
It can be shown that the decay of such a state results in
\begin{eqnarray}
  \tau_\mu \simeq \gamma\tau_\mu^0 
  \left[ 1+{2\over 3} (\lambdabar/{\cal L})^2\right],
\end{eqnarray}
so that the deviation from the hypothesis of locality is very small
($\sim 10^{-25}$) in this case but definitely nonzero.

\section{Discussion}
General relativity is a consistent theory of pointlike coincidences involving
classical point particles and rays of radiation.
The theory is robust and can be naturally extended to include wave phenomena 
(\emphq{minimal coupling});
however, general relativity is expected to have limited significance in this
regime.
From a basic standpoint, the main difficulty is the hypothesis of locality.

An accelerated observer passes through a continuous infinity of hypothetical 
inertial observers;
therefore, the most general linear connection between the field measured
by the accelerated observer $\cal{F}_{\alpha\beta}(\tau)$ and
the locally measured field
$F'_{\alpha\beta}(\tau)=F_{(\alpha)(\beta)} (\tau)$ that is consistent with 
causality is
\begin{equation}
  {\cal F}_{\alpha\beta}(\tau) = F'_{\alpha\beta}(\tau) +
  \int_0^\tau {\cal K}_{\alpha\beta}{}^{\gamma\delta}(\tau,\tau')\,
  F'_{\gamma\delta}(\tau') \,\d \tau'.
  \label{equ:33}
\end{equation}
Here the observer is inertial for $\tau\leq 0$ and the absence of the kernel
${\cal K}$ would be equivalent to the hypothesis of locality;
moreover, if ${\cal K}$ is directly connected with acceleration, then the 
deviation from the hypothesis of locality is generally of order 
$\lambdabar/{\cal L}$.
Assuming that ${\cal K}$ is a convolution-type kernel (i.e. it depends only
on $\tau-\tau'$), it is possible to determine ${\cal K}$ uniquely based 
on the assumption that no observer can ever stay at rest with respect to a
basic radiation field.
This is simply a generalization of the well-known result of Lorentz invariance,
so that the \emph{motion} of an electromagnetic wave would then become
independent of the observer.
We extend the observer independence of wave notion to all basic radiation
fields and elevate this notion to the status of a fundamental
physical principle \cite{ref28}.
Writing equation (\ref{equ:27}) as $F'=\Lambda F$, our basic assumption
implies that the \emph{resolvent} kernel ${\cal R}$ is given by
\cite{ref31}
\begin{equation}
  {\cal R} = {\d \Lambda(\tau)\over \d \tau} \Lambda^{-1}(0).
\end{equation}
It follows that for a scalar field ($\Lambda=1$),
${\cal R}=0$ and hence ${\cal K}=0$;
therefore, an observer can in principle stay at rest with respect to a scalar 
field.
This is contrary to our basic assumption, which then excludes fundamental 
scalar fields.
In this way, a nonlocal theory of accelerated observers has been developed that
is in agreement with all available observational data \cite{ref31}.
Moreover, novel inertial effects are predicted by the nonlocal theory.
For instance, let us recall the thought experiment (cf. Figure \ref{fig:2})
involving plane electromagnetic radiation of frequency $\omega$ normally 
incident on an observer rotating counterclockwise with 
$\Omega\ll\omega$;
the nonlocal theory predicts that the field amplitude measured by the observer
is larger by a factor of $1+\Omega/\omega$ for positive helicity radiation
and smaller by a factor of $1-\Omega/\omega$ for negative helicity radiation.
For radio waves with $\lambdabar\simeq 1\;\textrm{cm}$ and an observer
rotating at a frequency of 50 Hz, we have 
$\Omega/\omega=\lambdabar/{\cal L} \simeq 10^{-8}$.

Finally, it should be mentioned that no thermal ambience is encountered for
an accelerated observer on the basis of the approach adopted in this paper.
This is consistent with the absence of any experimental evidence for such
a thermal ambience at present \cite{ref20}.
That is, either (\ref{equ:27}) or (\ref{equ:33}) can be used to determine 
the quantum radiation field according to an accelerated observer once the
quantum field in the inertial frame is given.
Indeed, the nonlocal theory has been developed based on the assumption that 
no quanta are created or destroyed merely because an observer accelerates
(\emphq{quantum invariance condition}).

\end{document}